\documentclass[12pt]{article}
\usepackage{graphicx,amssymb, amsmath}
\newcommand{\mathsym}[1]{{}}

\input epsf.tex

\let\badcite=\cite
\def\cite{~\badcite}

\def\slashchar#1{\setbox0=\hbox{$#1$}           
   \dimen0=\wd0                                 
   \setbox1=\hbox{/} \dimen1=\wd1               
   \ifdim\dimen0>\dimen1                        
      \rlap{\hbox to \dimen0{\hfil/\hfil}}      
      #1                                        
   \else                                        
      \rlap{\hbox to \dimen1{\hfil$#1$\hfil}}   
      /                                         
   \fi}
    \def\slashword#1{\setbox0=\hbox{$#1$}        
  \dimen0=\wd0                                   
   \setbox1=\hbox{/} \dimen1=\wd1                
   \ifdim\dimen0>\dimen1                         
      \rlap{\hbox to \dimen0{\hfil\bf---\hfil}} %
      #1                                         %
   \else                                         
      \rlap{\hbox to \dimen1{\hfil$#1$\hfil}}    
      /                                          
    \fi}                                         %

\catcode`@=11
\newdimen\vbigd@men                             

\def\vbig#1#2{{\vbigd@men=#2\divide\vbigd@men by 2%
   \hbox{$\left#1\vbox to \vbigd@men{}\right.\n@space$}}}
\catcode`@=12

\catcode`@=11
\def\citenum#1{\csname b@#1\endcsname}
\catcode`@=12

\begin{document}
\begin{titlepage}

\begin{flushright}
{SCUPHY-TH-08006}\\
{CAS-KITPC/ITP-085}\\
\end{flushright}

\bigskip\bigskip

\begin{center}{\Large\bf\boldmath
A Simple Mass Reconstruction Technique for SUSY particles at the LHC }
\end{center}
\bigskip
\centerline{\bf N. Kersting\footnote{nkersting@scu.edu.cn}}
\centerline{{\it Physics Department, Sichuan University }}
\centerline{{\it Chengdu, P.R. China 610065}}
\centerline{{\it and}}
\centerline{{\it Kavli Institute for Theoretical Physics China, CAS }}
\centerline{{\it  Beijing, P.R. China 100190}}
\bigskip

\begin{abstract}

 It is often true that an invariant mass constructed from visible decay products of a heavy particle may attain a maximum(or minimum) for a certain kinematic configuration only --- this fact can be used to reconstruct relevant particle masses from observed decay product momenta of events near the invariant mass endpoint. MSSM neutralino and chargino mass reconstruction at the LHC from multi-lepton endstates is illustrated by way of example.

\end{abstract}

\newpage
\pagestyle{empty}

\end{titlepage}


\section{Introduction}

Within the first few years of LHC data collection and analysis, a key issue will not only be the search for general signs of New Physics (NP)  beyond the Standard Model (SM), but also quantitative measurement of any NP particles produced. NP mass spectra, in particular, can offer an important handle on discriminating between different NP models. In the highly theoretically-motivated Minimal Supersymmetric(SUSY) Standard Model (MSSM), for example, many of the 100+ free input parameters, among which various relations are predicted by specific models of SUSY-breaking, are directly coupled to values of SUSY masses. In contrast to the situation at a lepton collider, where the center-of-mass (CM) collision energy can be precisely tuned to sweep through NP mass resonances, the LHC produces partonic collisions whose CM energies vary unpredictably from event to event, washing out potential resonance structures. Moreover, NP states may decay only partially to visible and detectable particles, as in the R-parity-conserving MSSM, where sparticles typically cascade down to SM states plus an even number of invisible Lightest SUSY Particles (LSPs). Any MSSM mass reconstruction technique applied to LHC data cannot therefore depend on the precise CM energy or total visibility of decay products.

In answer to these demands, the phenomenological community has innovated a number of different mass reconstruction methods, the most standard and well-tested among these relying on measuring endpoints of various invariant mass distributions constructed from visible final leptonic and/or jet 4-momenta\cite{invmass}. In the MSSM, for example, 3-body decays of the second neutralino to a pair of leptons plus the LSP, $\widetilde{\chi}_{2}^0 \to   \ell^+ \ell^-  \widetilde{\chi}_{1}^0$, gives rise to a dilepton invariant mass distribution $M_{\ell^+ \ell^-}$ which cuts off relatively sharply at $M_{\ell^+ \ell^-}^{max} = m_{\widetilde{\chi}_{2}^0} - m_{\widetilde{\chi}_{1}^0}$. With a sufficiently large event sample this mass difference can be determined very precisely (to the sub-GeV level), but the values of the individual masses themselves are undetermined --- this is typical of the endpoint method, where the endpoint is generally some function of a number of NP masses.

The present work introduces the idea that there can be more information at the endpoint than just its numerical value --- events lying at the endpoint may arise from a unique kinematic configuration of final momenta in a decaying particle's frame --- and
additional analysis  can be done to find the masses. Let this be defined as the DK (Decay-frame Kinematics) technique. As a first example,
 consider the 3-body decay noted above\footnote{The case of 2-body decays is quite similar and will be reserved for a future work\cite{Kersting:wip}.}. An event at the dilepton endpoint must be such that, in the $\widetilde{\chi}_{2}^0$ decay frame, the $\widetilde{\chi}_{1}^0$ is at rest while the leptons are produced with equal and opposite momenta (the dilepton system has zero velocity). This fact allows us to find the velocity of
the $\widetilde{\chi}_{2}^0$ in the detector frame: namely, it is the observed velocity of the dilepton system, $(\overrightarrow{p}_{\ell^+}+\overrightarrow{p}_{\ell^-})/(E_{\ell^+} +E_{\ell^-} )$.
When we apply the corresponding Lorentz boost to the LSP 4-vector $(m_{\widetilde{\chi}_{1}^0}, 0, 0, 0)$ and match to observed missing transverse momenta, the LSP mass can be easily solved for --- twice in fact, from each transverse direction. The fact that R-parity requires a pair of such decays in each event changes nothing essential: one simply finds both $\widetilde{\chi}_{2}^0$ velocities from events at a \textit{double-endpoint}, i.e. where both dilepton invariants are maximal, matching missing momentum to the sum of LSP momenta. Although no event ever lies exactly at a double-endpoint, a large number of events may be within a tolerably small neighborhood of it --- one would expect that using these to reconstuct the LSP mass as above would give a distribution  peaked at the correct value.
There are, of course, immediate practical limitations  --- how small does the neighborhood have to be, do detector effects and backgrounds smear the peak beyond recognition, etc. --- which must ultimately fall to a Monte Carlo test in order to properly address.

As a second example of the DK technique in this paper, we shall consider  production of neutralino-chargino pairs  $\widetilde{\chi}_{2}^0 \widetilde{\chi}_{1}^\pm$ in the MSSM, with subsequent decays
$\widetilde{\chi}_{2}^0 \to
    \ell^+ \ell^- \widetilde{\chi}_{1}^0$ and $\widetilde{\chi}_{1}^\pm \to   \ell^\pm \nu \widetilde{\chi}_{1}^0$. Though slightly more complicated than the case of neutralino-pair decays,  the same principle works:
 events at a particular endpoint belong to a certain class of kinematics  in the  $\widetilde{\chi}_{2}^0$ and $\widetilde{\chi}_{1}^\pm$ decay frames, allowing us to find the individual velocities of these latter; the LSP 4-momenta can then be boosted and matched to observed missing momentum for only one (correct) value of the chargino mass.

 The structure of this work will be as follows: Section \ref{sec:zz} will illustrate the DK technique in more detail in the case of neutralino-pair production, where the goal is LSP mass-reconstruction --- this includes a Monte Carlo simulation to test how well the technique might actually work with LHC data. Section \ref{sec:wz} then does the same for neutralino-chargino pair production, again seeing how well the chargino mass can be reconstructed. Section \ref{sec:conc} summarizes these results, comments on the relation to other mass reconstruction techniques in the literature, and discusses general applicability.

\section{Neutralino-Neutralino Modes}
\label{sec:zz}
\subsection{Theory}
Consider production of neutralino pairs in the MSSM which undergo 3-body decays to electrons and muons:
\begin{equation} \label{zizjdecay2}
pp \to \mathbb{X}\to \mathbb{X}' + \widetilde{\chi}_{i}^0(\to
    e^+ e^- \widetilde{\chi}_{1}^0)~ \widetilde{\chi}_{j}^0 (\to   \mu^+ \mu^-  \widetilde{\chi}_{1}^0)
\end{equation}
 where $\mathbb{X}$ represents either Z* or any MSSM production channel from a Higgs ($H^0$ or $A^0$) or colored gluino/squark cascades, while $\mathbb{X}'$ are SM states potentially produced in association, all irrelevant to the current discussion; note we could include  $e^+ e^- e^+ e^-$ and $\mu^+ \mu^- \mu^+ \mu^-$ endstates as well, but at the price of a two-fold ambiguity in lepton pairing in what follows.
Physical observables from one event thus consist of four leptonic 4-momenta $p_{e^\pm,\mu^\pm}$ (from which we may construct the usual dilepton invariant masses, $M_{ee}$ and $M_{\mu\mu}$) and missing momentum in two transverse directions, assumed equal to the sum of the two $\widetilde{\chi}_{1}^0$s' transverse momenta, $p_{1,2}^T$. Decay kinematics (see Appendix) allow us to write the following list of constraints on the relevant neutralino masses (hereafter we abbreviate $m_i \equiv
m_{\widetilde{\chi}_{i}^0}$):
\begin{eqnarray}
\label{constraint1}
  |\overrightarrow{p}_{e^+}' + \overrightarrow{p}_{e^-}'| = \frac{1}{2 m_i}\sqrt{(M_{ee}^2 - m_1^2-m_i^2)^2 - 4 m_1^2 m_i^2}\\
  |\overrightarrow{p}_{\mu^+}' + \overrightarrow{p}_{\mu^-}'| = \frac{1}{2 m_j}\sqrt{(M_{\mu\mu}^2 - m_1^2-m_j^2)^2 - 4 m_1^2 m_j^2}  \\ \label{constraint3}
   (\overrightarrow{p}_1 + \overrightarrow{p}_2)^T = \slashchar{\overrightarrow{p}}^T ~\mathrm{(observed)}
\end{eqnarray}
where leptonic momenta are written in the frame of the respective parent neutralino, i.e.
${p}_{e^\pm}' = \mathbf{\Lambda}_1 {p}_{e^\pm}$ and ${p}_{\mu^\pm}' = \mathbf{\Lambda}_2 {p}_{\mu^\pm}$, defining the appropriate Lorentz transformations $\mathbf{\Lambda}_{1,2}$.
In general, of course, we do not know the Lorentz boosts $\overrightarrow{\beta}_{1,2}$ from which $\mathbf{\Lambda}_{1,2}$ are constructed, so (\ref{constraint1})-(\ref{constraint3}) represents  a system of four equations (or six, if the dilepton endpoints $M_{\ell \ell}^{max}= m_{i,j} - m_1$ are known) for nine unknowns ($\overrightarrow{\beta}_{1,2}$ and the masses $m_{1,i,j}$) which obviously cannot be solved uniquely for the masses.

\begin{figure}[!htb]
\begin{center}
\includegraphics[width=2.5in]{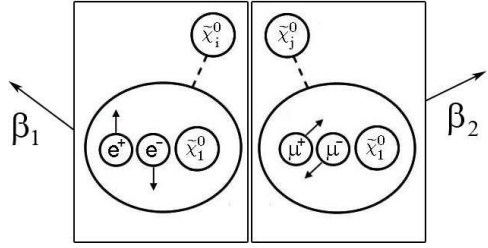}
\end{center}
\caption{\small \emph{ An event with maximal $M_{ee}$ and $M_{\mu\mu}$: though the decaying neutralinos $\widetilde{\chi}_{i,j}^0$ may be moving with any velocity
$\beta_{1,2}$ in the lab frame, in each respective decay frame the leptons have equal and opposite momenta while the
$\widetilde{\chi}_{1}^0$ is at rest.}
}
 \label{fig:zizj}
\end{figure}
Supposing, however, we have an event where the invariant masses $M_{ee}$ and $M_{\mu\mu}$ are maximal, as in Fig.~\ref{fig:zizj}, an enlarged system of constraints results:
\begin{eqnarray}
\label{constraint4}
 M_{ee} = m_i - m_1 \\ \label{constraint5}
 M_{\mu \mu} = m_j - m_1 \\ \label{constraint6a}
  |\overrightarrow{p}_{e^+}' + \overrightarrow{p}_{e^-}'| = 0 \\ \label{constraint6b}
  |\overrightarrow{p}_{\mu^+}' + \overrightarrow{p}_{\mu^-}'| = 0 \\ \label{constraint7}
   (\overrightarrow{p}_1 + \overrightarrow{p}_2)^T = \slashchar{\overrightarrow{p}}^T ~\mathrm{(observed)}
\end{eqnarray}
which now gives ten equations ((\ref{constraint6a}) and (\ref{constraint6b}) stand for three constraints each) for the nine unknowns\footnote{Strictly speaking, there is a two-fold ambiguity between (\ref{constraint4}) and (\ref{constraint5}), unless $i=j$.}, allowing us to actually overconstrain the masses $m_{1,i,j}$.
The $\overrightarrow{\beta}_{1,2}$ which satisfy (\ref{constraint6a}) and (\ref{constraint6b}), making the total momentum of each lepton pair zero, are uniquely given by
\begin{equation}\label{betaeqn1}
    \overrightarrow{\beta}_1 = \frac{\overrightarrow{p}_{e^+} + \overrightarrow{p}_{e^-}}{E_{e^+} + E_{e^-}} ~~~~~~~~
   \overrightarrow{\beta}_2 = \frac{\overrightarrow{p}_{\mu^+} + \overrightarrow{p}_{\mu^-}}{E_{\mu^+} + E_{\mu^-}}
\end{equation}
 Now the corresponding $\mathbf{\Lambda}_{1,2}$ which take the $e^+ e^- \widetilde{\chi}_{1}^0$ and $\mu^+ \mu^- \widetilde{\chi}_{1}^0$  systems to their respective $\widetilde{\chi}_{i,j}^0$-rest frames also bring each $\widetilde{\chi}_{1}^0$ to rest (a condition of $M_{ee}$ and $M_{\mu\mu}$ being maximal): their 4-momenta in these frames must thus be $(m_1,\overrightarrow{0})$, which, when inverse-Lorentz-transformed by $\mathbf{\Lambda}_{1,2}^{-1}$ to give
  $(m_1 \gamma_{1,2}~,~ m_1 (\overrightarrow{\beta} \gamma)_{1,2} )$,
  must agree with the observed missing momentum $\slashchar{\overrightarrow{p}}^T$; a matching condition along each transverse direction (say $\hat{x}$ and $\hat{y}$) then gives two independent determinations of $m_1$:
\begin{equation} \label{m1eqn}
  m_1' = \frac{\slashchar{\overrightarrow{p}}_x}{(\beta_x \gamma)_1 + (\beta_x \gamma)_2} ~~~~~~~~~~~ m_1'' = \frac{\slashchar{\overrightarrow{p}}_y}{(\beta_y \gamma)_1 + (\beta_y \gamma)_2}
\end{equation}
Since we are assuming that  both $M_{ee}$ and $M_{\mu\mu}$  are precisely maximal (the perfect event of Fig.~\ref{fig:zizj}), we would of course get $m_1' = m_1''= m_1$; in practice, of course, we can only expect to capture an event within some neighborhood $\epsilon$ of the endpoints, $M_{ee,\mu\mu}= M_{ee, \mu \mu}^{max} \pm \epsilon$, in which case one can show (see Appendix) that $m_1'$ and $m_1''$ are approximately $ \sim m_1 \pm \sqrt{2\epsilon m_1}$. One might then expect that applying (\ref{betaeqn1}) and (\ref{m1eqn}) to a sample of events near the endpoint should give a distribution of $m_1'$ and $m_1''$ peaked near $m_1$ with a $O(\sqrt{2\epsilon m_1})$ spread.

 As a technical caveat to the above, things could go wrong if $\slashchar{\overrightarrow{p}}^T=0$  : this occurs if the neutralinos $\widetilde{\chi}_{i,j}^0$ happen to travel in opposite transverse directions with precisely the correct velocities, i.e. $(\beta_{x,y} \gamma)_1 + (\beta_{x,y} \gamma)_2$ = 0. In this case the
 formulae (\ref{m1eqn}) give us indeterminate solutions $m_1' = m_1'' = 0/0$. Even if $\slashchar{\overrightarrow{p}}^T$ is finite but small (say a few GeV) we might still worry that statistical fluctuations to $\slashchar{\overrightarrow{p}}^T \approx 0$ would likewise throw off the solutions (\ref{m1eqn}).
This will not be a serious concern, however, if most events contain significant ``upstream transverse momentum" (UTM)\cite{Barr} in $\mathbb{X}$ or $\mathbb{X}'$, and
 we will presently see that a Monte Carlo simulation of a typical LHC environment,
  adding in experimental effects such as measurement errors and inherent finiteness of detector resolution, as well as the inclusion of SM and MSSM backgrounds, basically confirms the robustness of DK mass reconstruction.

\subsection{Monte Carlo Test}

In order to see how well the above programme might work with real data, let us apply it to Monte Carlo simulated LHC data, choosing for definiteness the following MSSM parameter point:
\begin{eqnarray*}
  \mu = 390\, \hbox{GeV}  & tan \beta = 10 & m_A = 400 \, \hbox{GeV}   \\
   M_1 = 100\, \hbox{GeV} &  M_2 = 123\, \hbox{GeV}  & m_{\tilde{g}} = 605\, \hbox{GeV} \\
  m_{\tilde{q}} = 500\, \hbox{GeV} & m_{\widetilde{\ell_L}, \tilde{\tau}} = 300\, \hbox{GeV} & m_{\widetilde{\ell_R}} = 130\, \hbox{GeV}
\end{eqnarray*}
which is a modification of the SPS1a' benchmark point\cite{benchmarks} enhancing the $\widetilde{\chi}_{2}^0 \to \ell^+ \ell^- \widetilde{\chi}_{1}^0$ branching ratio.

Colored sparticles, dominating the bulk of the inclusive SUSY cross section ($\sim 70~pb$) at this point,
cascade predominantly to $\widetilde{\chi}_{2}^0$ ($\sim 30\%$) or $\widetilde{\chi}^\pm_1$ ($\sim 60\%$), excepting R-handed squarks (which mostly decay directly to the LSP). As can be seen from the gaugino mass spectrum in Table \ref{tab:mass}, the mass splitting between $\widetilde{\chi}_{2}^0$ and $\widetilde{\chi}_{1}^0$ (the LSP) of
$m_2 - m_1 = 18.7\, \hbox{GeV}$ is too small to allow anything but $\widetilde{\chi}_{2}^0$ decay through an off-shell intermediary (slepton or Z*) as desired: $\widetilde{\chi}_{2}^0 \to \ell^+ \ell^- \widetilde{\chi}_{1}^0$  has $BR \approx 90\%$ (next of importance is $\widetilde{\chi}_{2}^0 \to q q' \widetilde{\chi}_{1}^0$ with $BR \approx 9\%$). SUSY $e^+ e^- \mu^+ \mu^-$ endstates should therefore arise almost entirely from
 $\mathbb{X}\to \mathbb{X}' + \widetilde{\chi}_{2}^0(\to
    e^+ e^- \widetilde{\chi}_{1}^0)~ \widetilde{\chi}_{2}^0 (\to   \mu^+ \mu^-  \widetilde{\chi}_{1}^0)$.
\begin{table}
 \caption{\small \emph{Gaugino masses (in GeV) at the MSSM parameter point considered.}}
    \begin{center}
     \begin{tabular}{|c|c|c|c|c|c|} \hline
   ${\widetilde\chi}^0_1$
 &  ${\widetilde\chi}^0_2$
 &  ${\widetilde\chi}^0_3$
 &  ${\widetilde\chi}^0_4$
 &  ${\widetilde\chi}^\pm_1$
 &  ${\widetilde\chi}^\pm_2$
 \\
 \hline
 $96.5$  & $115.2$ & $393.8$ & $404.9$ & $114.2$ & $406.0$ \\ \hline
       \end{tabular}
    \end{center}
 \label{tab:mass}
\end{table}

 Proton-proton collisions corresponding to $100~fb^{-1}$ of LHC luminosity are then generated with HERWIG~6.5\cite{HERWIG65} (coupled to the CTEQ6 parton distribution functions\cite{CTEQ6}) for all relevant SUSY processes ($pp \to$ any pair of
  $\{\tilde{q}, \tilde{g}, \widetilde{\chi}^\pm, \widetilde{\chi}^0 \}$, $pp \to \tilde{\ell}\tilde{\ell}$) and SM backgrounds (for a hard 4-lepton signal only $Z^{(*)}Z$ is sizeable\cite{Bian}), coupled to a simplified detector simulator\footnote{An identical set-up was employed in the author's previous publications\cite{Kersting:ss, Huang1} and the reader is referred
there for additional details.}  which only passes events with four hard, isolated\footnote{No tracks of other charged particles are present in a $r = 0.3\, \hbox{rad}$ cone around the lepton, with less than $3\, \hbox{GeV}$ of energy deposited into the
electromagnetic calorimeter for $0.05\, \hbox{rad} < r < 0.3\, \hbox{rad}$, and $p_T^\ell > 10, 8\, \hbox{GeV}$ for $\ell = e,\mu$.}
leptons with flavor structure $e^+ e^- \mu^+ \mu^-$.
\begin{figure}[!htb]
\begin{center}
\includegraphics[width=2.5 in]{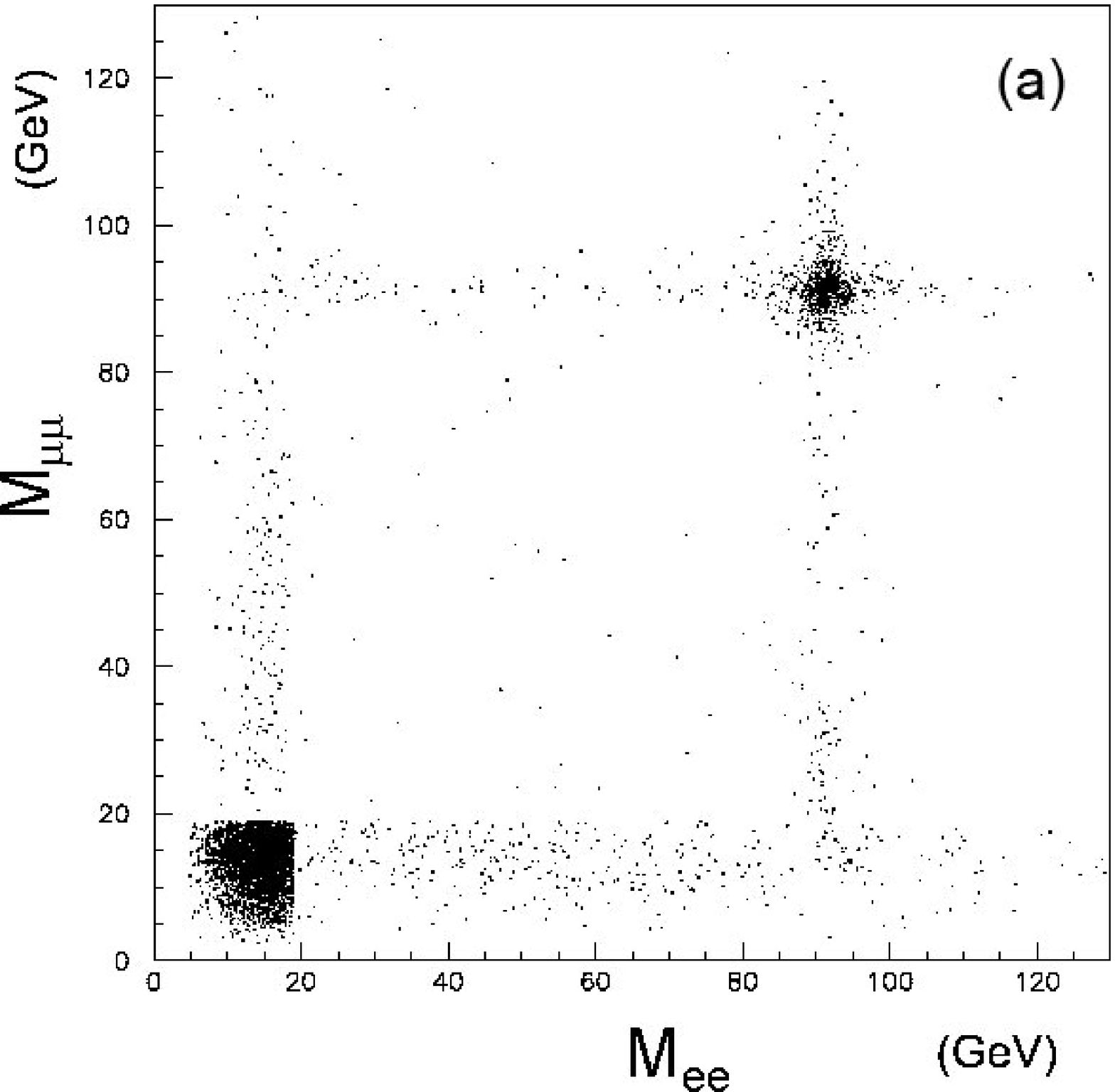}
\includegraphics[width=2.5 in]{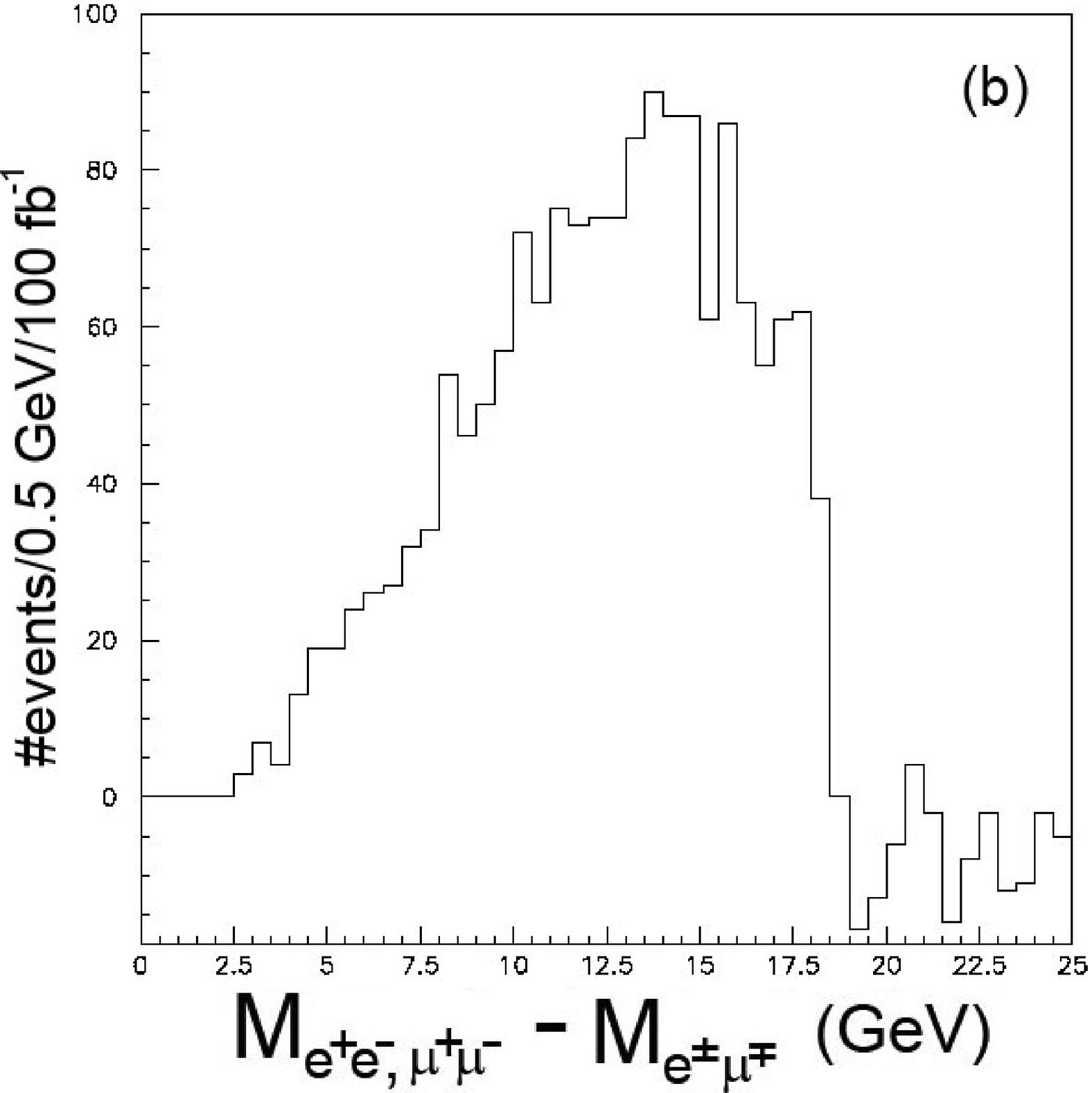}
\includegraphics[width=2.5 in]{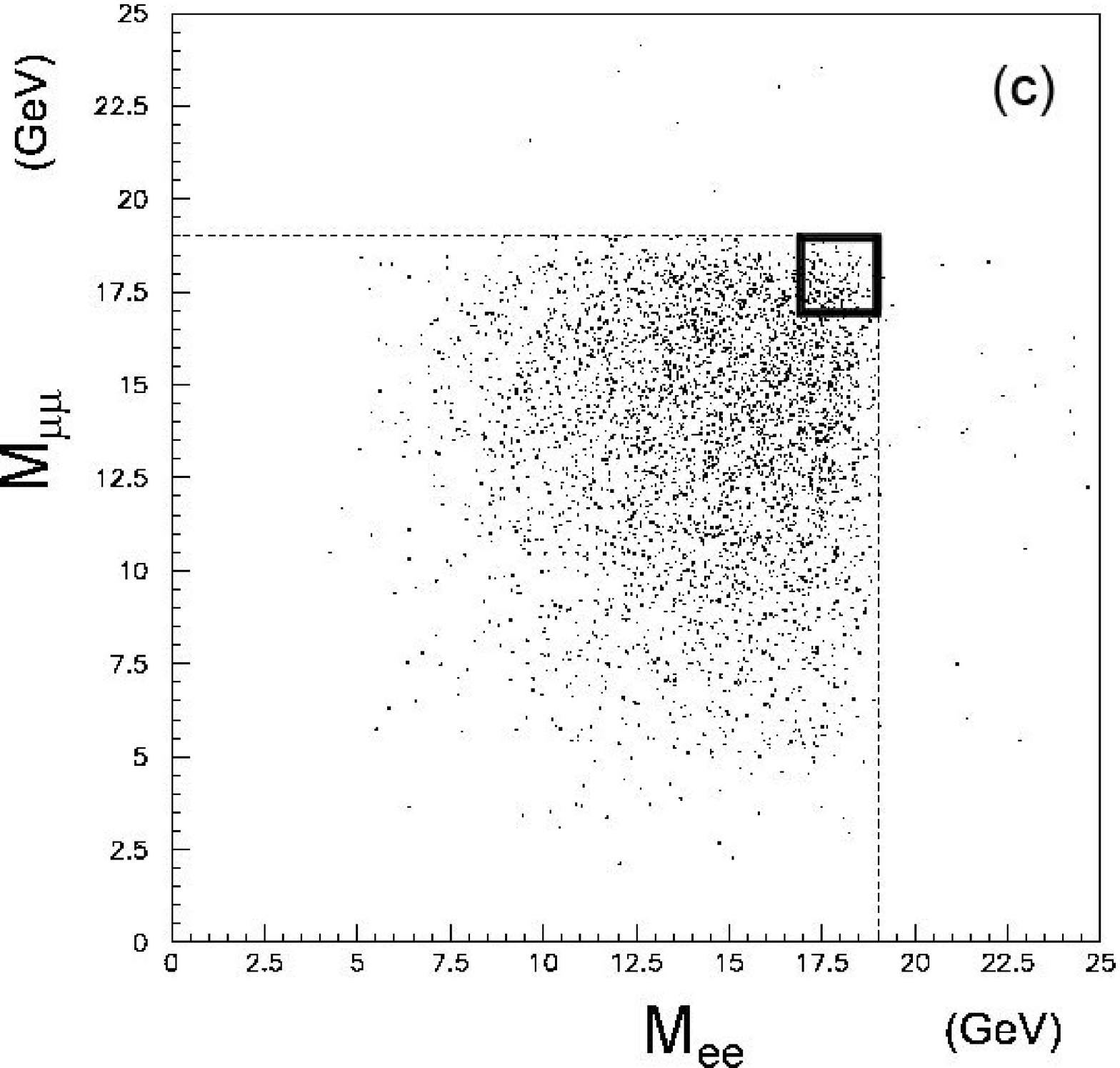}
\includegraphics[width=2.5 in]{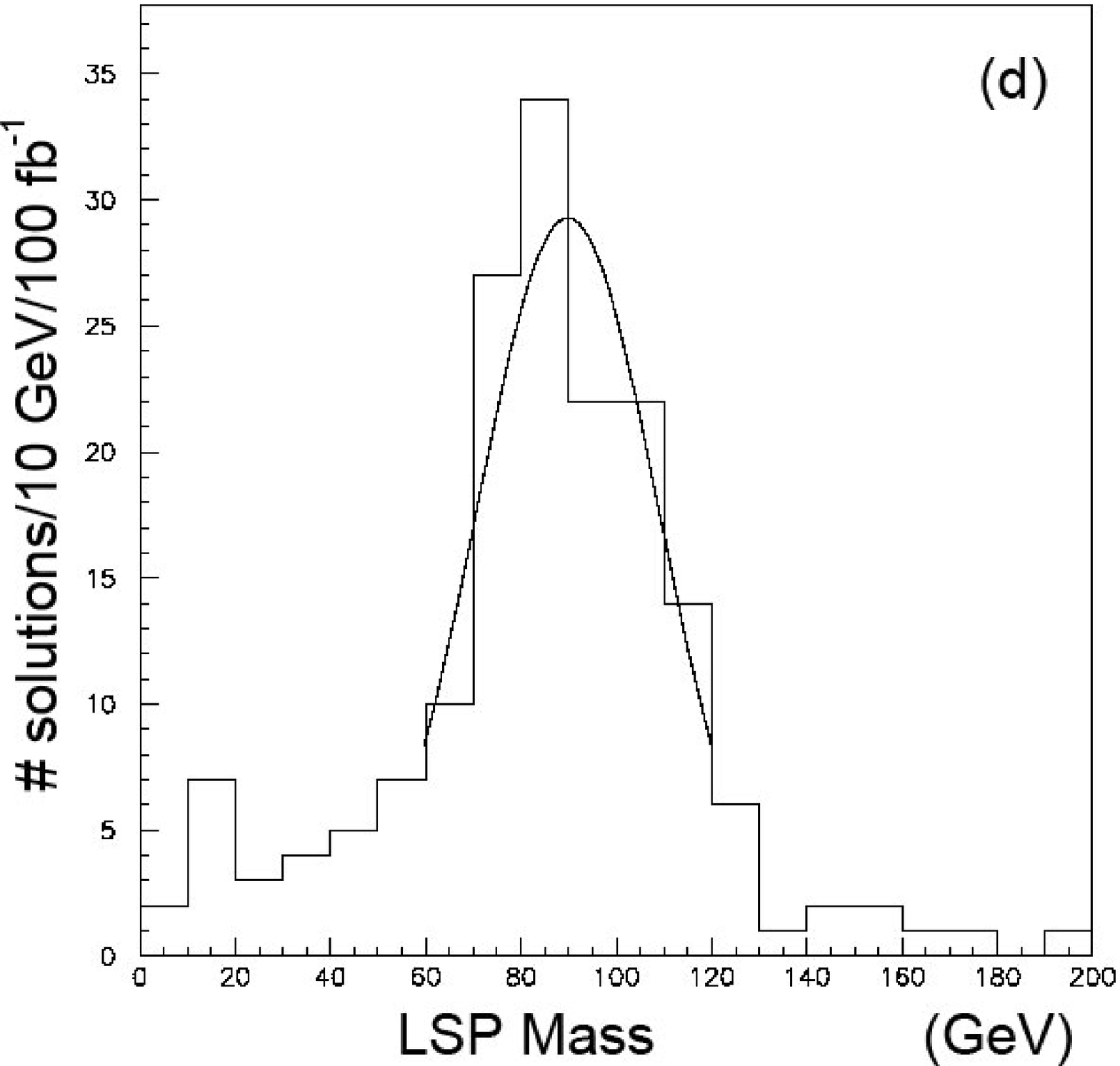}
\end{center}
\caption{\small \emph{ (a) Wedgebox plot ($100~fb^{-1}$) showing a very dense box with diffuse backgrounds  (except near the Z-pole).
 (b) Flavor-subtracted dilepton events give a sharp distribution with an edge of approximately $M_{\ell^+ \ell^-}^{max} \approx 19\,\hbox{GeV}$. (c) Enlarged view of the wedgebox within this edge; events for analysis are taken from the small boxed region of area $2\times 2\,\hbox{GeV}^2$ shown. (d) Distribution of reconstructed LSP mass, with Gaussian fit overlaid: $m_{\widetilde{\chi}_{1}^0} = 89.9 \pm 19.1\,\hbox{GeV}$.
}
}
 \label{fig:neut}
\end{figure}
The wedgebox plot of Fig.~\ref{fig:neut}a shows all surviving events ($\sim 5000$ of these) by position in ($M_{ee}$,$M_{\mu\mu}$)-space and is well-suited to picking out those arising from the desired $\widetilde{\chi}_{i}^0 \widetilde{\chi}_{j}^0$-origin\cite{Bisset}; in this case the dense box-like structure in the corner
 ($M_{ee,\mu \mu} < 20\, \hbox{GeV}$) arises from $\widetilde{\chi}_{2}^0 \widetilde{\chi}_{2}^0$-pair decays as expected; areas of the wedgebox plot not relevant to the present study include the `wings' projecting along either axis, arising from decays of more sparsely-produced $\widetilde{\chi}_{2}^0 \widetilde{\chi}_{3,4}^0$ and $\tilde{\ell}\tilde{\ell}$ pairs, as well as the two lines of points concentrated around $M_{ee,\mu \mu} \approx 90\, \hbox{GeV}$ due to the $Z^{(*)}Z$ background.

 Focusing on the corner region then, a flavor-subtracted dilepton invariant mass distribution (Fig.~\ref{fig:neut}b) constructed from hard 2-lepton events (over 27000 of these) is most useful for pinpointing the endpoint: for our purposes it is enough to roughly identify $M_{\ell^+ \ell^-}^{max} \approx 19\,\hbox{GeV}$ by visual inspection (this is already in excellent agreement with the expected value of $m_2 - m_1 = 18.7\, \hbox{GeV}$). Note the less-than-triangular shape of this distribution does indicate the 3-body  nature of the $\widetilde{\chi}_{2}^0$-decays responsible. The $\widetilde{\chi}_{2}^0 \widetilde{\chi}_{2}^0$ box on the wedgebox plot is of course defined where
 $M_{ee,\mu \mu} < M_{\ell^+ \ell^-}^{max}$, and an enlarged view of this region (Fig.~\ref{fig:neut}c) allows us to select an event sample in the neighborhood of $M_{ee,\mu \mu} \sim M_{\ell^+ \ell^-}^{max}$ --- here `neighborhood' might be defined as $M_{\ell^+ \ell^-}^{max} - \epsilon < M_{ee,\mu \mu} < M_{\ell^+ \ell^-}^{max}$ for some optimal $\epsilon$, i.e. $\epsilon$ must be large enough to get more than a few events, yet not so large that the kinematic configuration of Fig.~\ref{fig:zizj} becomes a bad approximation for most events; at the present point $\epsilon = 2\, \hbox{GeV}$ seems best (see below), giving an event sample of size $O(10^2)$.
 For each such event, (\ref{betaeqn1}) and (\ref{m1eqn}) are applied from the measured leptonic momenta and missing transverse momenta, both $m'$ and $m''$ being added to the final distribution (Fig.~\ref{fig:neut}d) only if they agree with each other to within $20 \%$ --- this is very effective at eliminating not only background events which by chance might fall in the sampling region, but also legitimate $\widetilde{\chi}_{2}^0 \widetilde{\chi}_{2}^0$ events for which the DK technique would fail (mostly because of momentum mis-measurement effects in the detectors but also possibly from wrongly-assumed kinematics of Fig.~\ref{fig:zizj}) ---
  nevertheless, some events give consistent solutions which lie far from the prominent peak at $ 80-100\, \hbox{GeV}$. At any rate, a Gaussian fit\footnote{For lack of a known theoretical shape. Away from the peak the distribution is clearly non-Gaussian, a feature which the author has found true at other parameter points tested.} to the peak indicates a reconstructed LSP mass of $m_1 = 89.9 \pm 19.1\,\hbox{GeV}$, in safe agreement with the nominal value in Table~\ref{tab:mass}. The slight downwards bias in the central value seems to be a systematic effect, for increasing $\epsilon$ enhances the bias --- indeed, at lower $\epsilon$ (selecting a sample of events more closely approximating the kinematics of Fig.~\ref{fig:zizj}) the central value more closely approaches the nominal one, but at the price of statistics, the balancing of these two effects leading to the quoted value of $\epsilon = 2\, \hbox{GeV}$. Recall from the theoretical discussion above that the expected width of the LSP-mass distribution would then be at least $O(\sqrt{2\epsilon m_1}) = O(\sqrt{(2)(2)(90)}) \,\hbox{GeV} \approx 19\,\hbox{GeV}$, in good agreement with what is seen; a higher luminosity sample (even at full design luminosity of $300~fb^{-1}$) might permit a smaller optimal $\epsilon$, say $\epsilon \sim 1 \,\hbox{GeV}$, but  the distribution width of $\sqrt{2\epsilon m_1}$ would not be expected to decrease significantly --- application of DK therefore reliably locates the LSP mass, but only to a limited precision (probably $10 \%$ at best). This may, nevertheless, be entirely sufficient for discriminating among different SUSY models, and, as we shall see next, is also good enough to find the mass of the lighter chargino.

\section{Neutralino-Chargino Modes}
\label{sec:wz}

The foregoing has demonstrated how the DK technique reconstructs the LSP mass from 4-lepton endstates of neutralino pair ($\widetilde{\chi}_{2}^0 \widetilde{\chi}_{2}^0$) decays; let us now see how DK also reconstructs the chargino mass from 3-lepton endstates of neutralino-chargino ($\widetilde{\chi}_{2}^0 \widetilde{\chi}_{1}^\pm $) decays, again taking these to be through off-shell intermediates, i.e.
\begin{equation} \label{zizjdecay2}
pp \to \mathbb{X}\to \mathbb{X}' + \widetilde{\chi}_{2}^0(\to
    e^+ e^- \widetilde{\chi}_{1}^0)~ \widetilde{\chi}_{1}^\pm (\to   \mu^\pm \nu \widetilde{\chi}_{1}^0)
\end{equation}
or with $e \leftrightarrow\mu$; again, same-flavor endstates such as $e^+ e^- e^\pm$ or $\mu^+ \mu^- \mu^\pm$ could also be included in this analysis at the price of a two-fold ambiguity in lepton-pairing.

\begin{figure}[!htb]
\begin{center}
\includegraphics[width=2.5in]{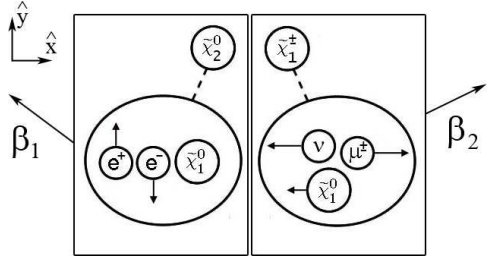}
\end{center}
\caption{\small \emph{ Events of interest for reconstructing the chargino mass: in the frame where the $\widetilde{\chi}_{2}^0$ decays at rest, the $e^\pm$ are along some global coordinates $\pm \hat{y}$ with maximal energy; likewise in the $\widetilde{\chi}_{1}^\pm$ decay frame the muon is along $\pm \hat{x}$ with maximal energy. The velocities $\beta_{1,2}$ are a priori unknown.}
}
 \label{fig:w1z2}
\end{figure}

With the $\widetilde{\chi}_{2}^0 \widetilde{\chi}_{2}^0$ modes of the last section, it was natural to classify 4-lepton events by their dilepton invariant masses  $M_{ee,\mu\mu}$ in the form of a wedgebox plot, events of interest being concentrated near
 $M_{ee} = M_{\mu\mu} =M_{\ell \ell}^{max}$. Here with 3-lepton events, a natural way to classify events is by the dilepton invariant $M_{ee}$ and the invariant $\overline{M}_{l2l}$, defined as
 \begin{equation}\label{ml2l}
     \overline{M}_{{l2l}}^4 \equiv  \{
  (p_{e^+} + p_{{e^-}}- p_{\mu^\pm})^4 +
  (p_{\mu^\pm} + p_{{e^+}}- p_{e^-})^4 +
  (p_{{e^-}}+ p_{\mu^\pm}-p_{e^+})^4  \}/3
 \end{equation}
since, as found in\cite{Kersting:ss}, events with maximal  $M_{ee}  = M_{\ell \ell}^{max}$ and minimal $\overline{M}_{l2l}$ arise from events as shown in Fig.~\ref{fig:w1z2}, i.e. if the decaying $\widetilde{\chi}_{2}^0$ and $\widetilde{\chi}_{1}^\pm$ had no  motion
 then the $e^\pm$ would have equal and opposite momenta (along say $\pm \hat{y}$) while the muon would be emitted perpendicular to $\hat{y}$, say along $\hat{x}$, also with maximal kinetic energy.

 Just as before, the required Lorentz transformation $\mathbf{\Lambda}_1$ to get the $e^+ e^-$ pair in the frame of the decaying neutralino (${p}_{e^\pm}' = \mathbf{\Lambda}_1 {p}_{e^\pm}$) is determined by
 \begin{equation}\label{betaeqnc}
    \overrightarrow{\beta}_1 = \frac{\overrightarrow{p}_{e^+} + \overrightarrow{p}_{e^-}}{E_{e^+} + E_{e^-}}
\end{equation}
There is unfortunately no such simple formula for $\overrightarrow{\beta}_2$,
 but it is nevertheless uniquely defined by the following system of equations describing conservation of the total missing 4-momentum $\slashchar{p}^\mu$,
\begin{equation}\label{pteqn}
   \slashchar{p}^\mu = \mathbf{\Lambda}_1^{-1}\left(
  \begin{array}{c}
    m_1  \\
   \overrightarrow{0} \\
  \end{array}
\right)
+
 \mathbf{\Lambda}_2^{-1}\left(
  \begin{array}{c}
   m_\pm -  E_{\mu^\pm}'  \\
    -  \overrightarrow{p}_{\mu^\pm}' \\
  \end{array}
\right) ~~~~~~~~ \left[ \begin{array}{c}
\left(
  \begin{array}{c}
     E_{\mu^\pm}' \\
    \overrightarrow{p}_{\mu^\pm}'  \\
  \end{array}
\right) \equiv \mathbf{\Lambda}_2 \left(
  \begin{array}{c}
     E_{\mu^\pm}  \\
     \overrightarrow{p}_{\mu^\pm} \\
  \end{array}
\right) \\
m_\pm \equiv m_{\widetilde{\chi}_{1}^\pm} \\
\end{array}
 \right]
\end{equation}
of which the two transverse components $\slashchar{\overrightarrow{p}}^T$ are measurable,
in addition to the three kinematic constraints
\begin{equation}\label{muenergy}
    \overrightarrow{p}_{e^\pm}' \cdot \overrightarrow{p}_{\mu^\pm}' = 0
     ~~~~,~~~~~ E_{\mu^\pm}' = \frac{{m^2_\pm} - m_1^2  }{2 m_\pm}
     ~~~~,~~~~M_{e e} = m_2 - m_1
\end{equation}
which events of the form shown in Fig.~\ref{fig:w1z2} must obey.
Thus, (\ref{pteqn}) and (\ref{muenergy}) compose a system of five equations for the six unknowns
$\{ \overrightarrow{\beta}_2,$ $m_1,~ m_2,~m_\pm\}$.
Supposing $m_1$ is already determined (say from the last section), then we are less one unknown and the
chargino mass $m_\pm$ can found from one (perfect) event, in principle. The usual practical caveats apply to this claim, so again we will have to check via Monte Carlo.

\subsection{Monte Carlo Test}

\begin{figure}[!htb]
\begin{center}
\includegraphics[width=2.5 in]{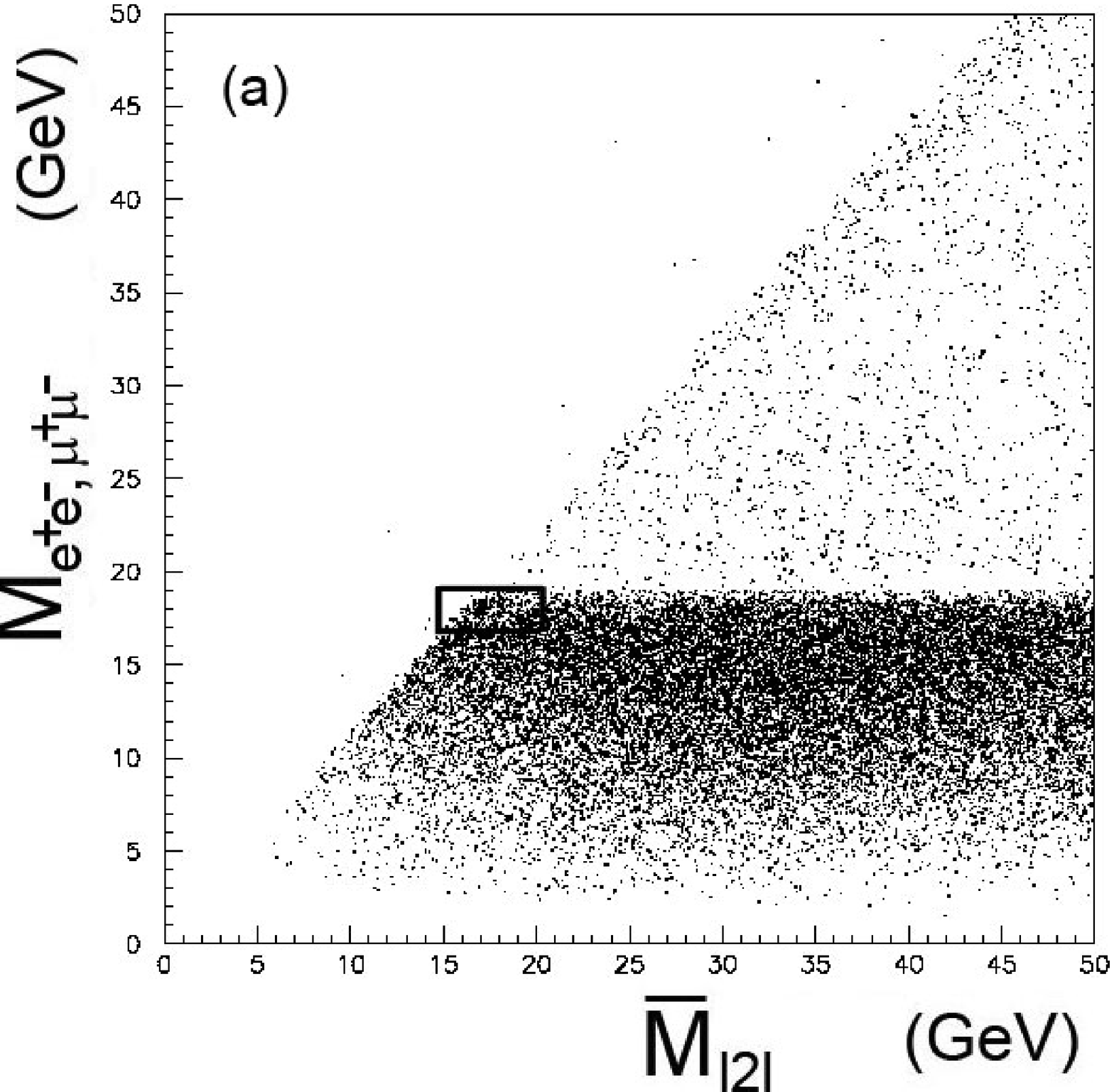}
\includegraphics[width=2.5 in]{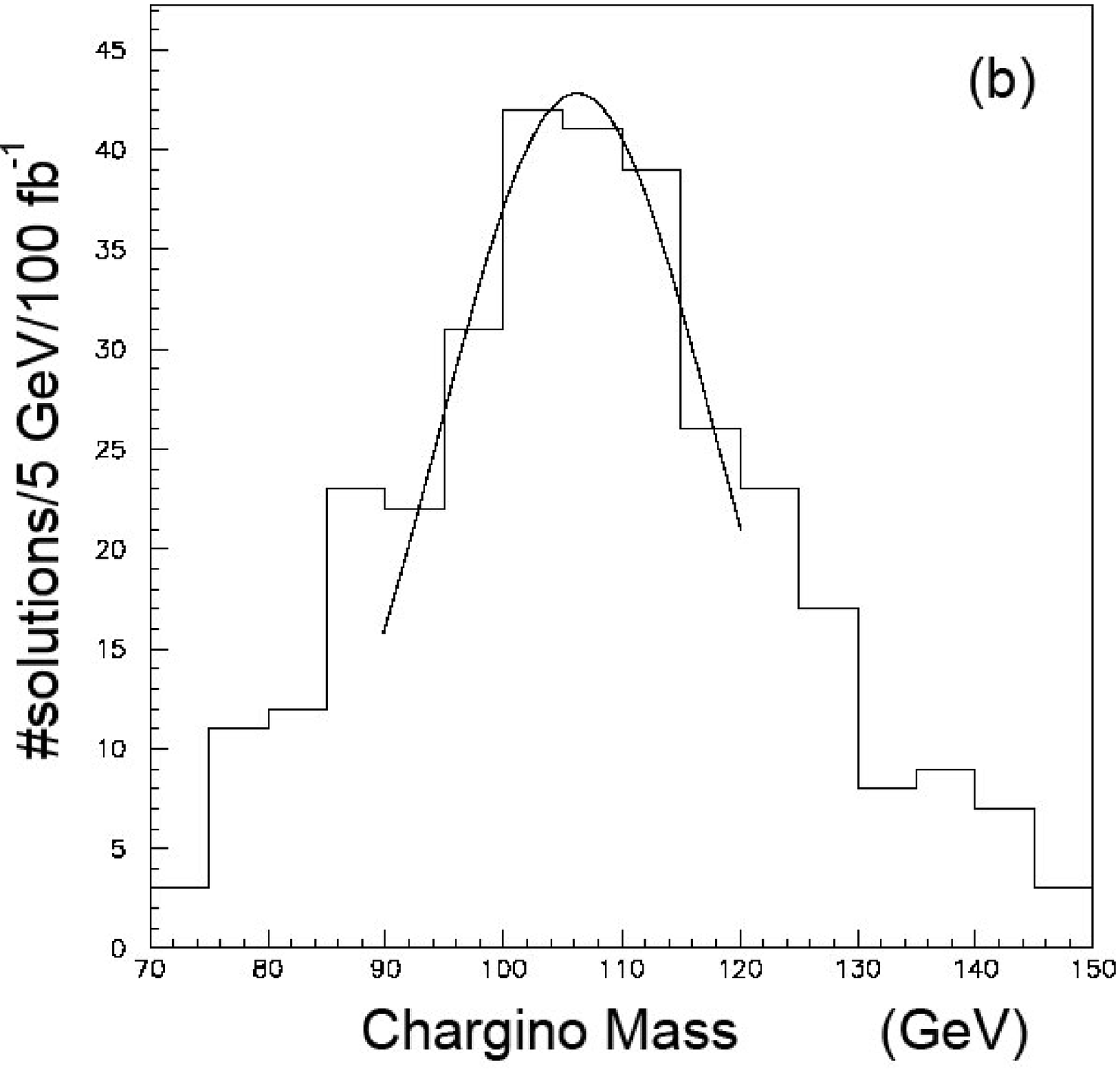}
\end{center}
\caption{\small \emph{ (a) Correlation between $M_{\ell^+ \ell^-}$ and $\overline{M}_{l2l}$ ($100~fb^{-1}$);
events are taken in the boxed region shown where $M_{\ell^+ \ell^-} = 18\pm 1\,\hbox{GeV}$ and
$\overline{M}_{l2l} < 20 \,\hbox{GeV}$.
(b) Distribution of reconstructed  chargino mass from events in this boxed region, assuming the reconstructed LSP mass of  $m_1 = 89.9 \pm 19.1\,\hbox{GeV}$. A Gaussian fit gives $m_{\widetilde{\chi}_{1}^\pm} = 106.2 \pm 11.6\,\hbox{GeV}$.}
}
 \label{fig:charg}
\end{figure}

Working with the same MSSM parameter point and  Monte Carlo setup as in the last section, signal and background processes (now including significant 3-lepton sources such as $W\gamma^{*}/Z$) for  $100~fb^{-1}$ integrated luminosity are generated and now filtered for events with only three hard and isolated leptons --- this, in addition to the fact that squark BR's to charginos  are typically twice those to neutralinos at this point, gives a much larger signal: the ``$M_{\ell^+ \ell^-}$ versus $\overline{M}_{l2l}$" plot of Fig.~\ref{fig:charg}a represents well over $4 \cdot 10^4$ events.
Though this plot gives us some discriminating power between signal and background, e.g. the majority of SM background events cluster about the line $M_{\ell^+ \ell^-} = m_Z$ (cut off by the plot), there is still a more-or-less uniform distribution of background events such as $\widetilde{\chi}_{2}^0 \widetilde{\chi}_{1}^\pm(\to \tau \nu \widetilde{\chi}_{1}^0)$ (the tau decaying leptonically) which will have to be tolerated.

From the previous theoretical discussion we know where to look on this plot, viz. the small rectangular box in Fig.~\ref{fig:charg}a containing events assumed to be of the type shown in Fig.~\ref{fig:w1z2}: by trial and error it is found that the several hundred events here satisfying $M_{\ell^+ \ell^-} = 18\pm 1\,\hbox{GeV}$ and
$\overline{M}_{l2l} < 20 \,\hbox{GeV}$ give an optimal sample in the sense discussed above in Section \ref{sec:zz}.
For each such event, $m_1$ is chosen in the interval $m_1 = 89.9 \pm 19.1\,\hbox{GeV}$ (flat priors) and the system of equations (\ref{betaeqnc})-(\ref{muenergy}) is solved for $m_\pm$. Only physical solutions ($m_\pm$ is real and $ > m_1$) are kept and plotted in Fig.~\ref{fig:charg}b. A Gaussian fit to the peak gives $m_\pm = 106.2 \pm 11.6\,\hbox{GeV}$, in good agreement with the actual value $m_\pm = 114.2\,\hbox{GeV}$; note again, however, the downwards bias in the central value. As with the fit to the LSP mass in the last Section, repeating the analysis with various smaller-sized sampling regions decreased the bias at the cost of statistics, so this is understood to be a systematic uncertainty of DK.

\section{Discussion and Summary}
\label{sec:conc}

We have now seen two specific examples
in the MSSM (from $\widetilde{\chi}_{2}^0 \widetilde{\chi}_{2}^0$ and $\widetilde{\chi}_{2}^0 \widetilde{\chi}_{1}^\pm$ pair decays)
of how DK operates: events near the endpoint of a particular invariant mass distribution (in the black boxes of Figs.~\ref{fig:neut}c and \ref{fig:charg}a, respectively)
can be used to reconstruct  masses ($m_1$ and $m_\pm$) to reasonable ($\sim 10-20\%$) precision, at least with the LHC luminosity and  MSSM input parameters assumed here.
 Though 3-body decays
$\widetilde{\chi}_{2}^0 \to   \ell^+ \ell^-  \widetilde{\chi}_{1}^0$ and  $\widetilde{\chi}_{1}^\pm \to   \mu^\pm \nu \widetilde{\chi}_{1}^0$ were assumed for the sake of simplicity, the DK technique also works (with minor adjustments) for 2-body decays through on-shell sleptons\cite{Kersting:wip}.

It is important to ask how this technique performs relative to other mass reconstruction methods available (excepting, of course, the traditional endpoint-value method, which is incorporated into DK).
First consider the reconstruction of neutralino masses.
There are now an array of methods which take advantage of the pair-production of neutralinos.
One class of ``Mass-Shell Techniques"(MST), represented in the work of\cite{Kawagoe} and\cite{Cheng},
essentially depends on maximizing the solvability of assumed mass-shell constraints in a given sample of events. This
seems quite effective for on-shell decays\footnote{But see\cite{Bisset} for some important caveats.}, but for the off-shell decay topologies  in the present work these methods
cannot be applied since there are not enough such constraints.
Recently fashionable ``transverse mass variable" methods\cite{mT2,mtvar}, e.g. $m_{T2}$, might be applied.
 It is, in fact, quite likely that such methods are closely related to DK in the sense that they also attach importance to events with particular kinematic configurations, e.g. the analysis in \cite{mtvar} would suggest that a $m_{T2}$ variable constructed for  $\widetilde{\chi}_{2}^0\widetilde{\chi}_{2}^0 \to   e^+ e^- \mu^+ \mu^- 2\widetilde{\chi}_{1}^0$  attains a well-defined (in terms of sparticle masses) maximum when the lepton pairs are either back-to-back or collinear and parallel; in one such development\cite{Barr}, for example,
 a ``constrained mass variable" $m_{2C}$ in events with large UTM proves quite powerful for
even a small number of events.  In any case, if any of the above methods are applicable, they should perform much better than DK alone, since all $\widetilde{\chi}_{2}^0 \widetilde{\chi}_{2}^0$ events (not just those close to a double-endpoint) are used.
As for decay modes with charginos, such as the $\widetilde{\chi}_{1}^\pm \widetilde{\chi}_{2}^0$ modes considered in this work, it would seem that other techniques would encounter difficulties
due to extra invisible particles (neutrinos) in the decay products. There is the interesting prospect, however, of somehow conjoining these techniques with DK for a better fit of the chargino mass.

\begin{figure}[!htb]
\begin{center}
\includegraphics[width=2.5in]{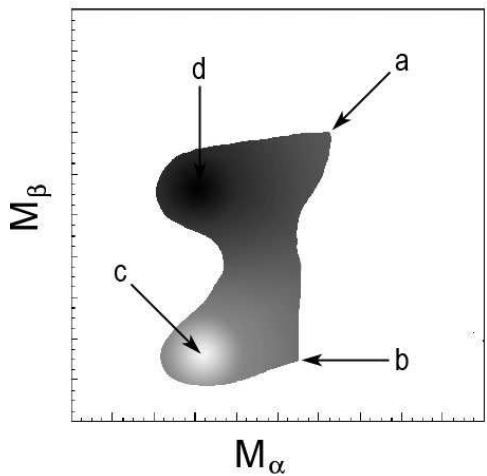}
\end{center}
\caption{\small \emph{General 2d Dalitz plot; (a)-(d) indicate special regions which may have
unique kinematics applicable to DK analysis. }
}
 \label{fig:gen}
\end{figure}

The DK technique is by no means limited to neutralino or chargino decays, nor just to MSSM applications.
Any NP decay chain which is subject to the traditional invariant mass endpoint method also can be analyzed with
DK.
The key is to choose invariant masses whose endpoints correspond to unique decay-frame kinematics,
e.g. $M_{\ell \ell}$(at its maximum) in the case of neutralino decays, or both $M_{\ell \ell}$ and $\overline{M}_{{l2l}}$
(which must be simultaneously maximal and minimal, respectively) for neutralino-chargino decays.
This statement  allows one to imagine how the
  DK technique might work in general:
  depending on the specific number and type of invariant masses one can construct from
  the endstates of a given decay chain, some d-dimensional Dalitz plot of these will be
  useful for isolating events with a unique decay-frame kinematic configuration.
Fig.~\ref{fig:gen} shows, for example, a hypothetical 2d Dalitz plot of some invariants
$M_\alpha$ and $M_\beta$: points which might exhibit unique, identifiable
decay-frame kinematics, hence amenable to DK analysis, include
(a)extrema\footnote{But beware: endpoints do not always arise from unique kinematic configurations; see\cite{Huang1,Gjelsten}.}, (b)kinks, (c)internal density minima or (d)maxima. Extension to
$d \ge 3$- dimensional Dalitz plots is feasible if enough events can be collected to fill in the shape.

\section*{Acknowledgements}
This work was funded in part by the Kavli Institute for Theoretical Physics (Beijing).

\newpage

\section*{Appendix}
\subsection*{Off-Shell Kinematics}
Consider the three-body decay $\widetilde{\chi}_{i}^0 \to
    \ell^+ \ell^- \widetilde{\chi}_{1}^0$ in the rest frame of the decaying neutralino. Working in the approximation where the lepton mass is zero, the leptonic four-momenta are $p^\mu_{\ell^\pm} = (E_\pm,~ E_\pm \hat{p}_\pm)$ and
    the dilepton invariant mass $M^2_{\ell^+ \ell^-} \equiv (E_+ + E_-)^2 - (E_+ \hat{p}_+ + E_- \hat{p}_- )^2$ can be
    rewritten as
    \begin{equation}\label{eq1}
        M^2_{\ell^+ \ell^-} = 2 E_+ E_- (1 - cos\theta)
    \end{equation}
    where $\theta$ is the angle between the leptons.
    Energy and momentum conservation further imply
    \begin{eqnarray} \label{eq2}
      E_+ + E_- + E_1 &=& m_i \\ \label{eq3}
      E_+ \hat{p}_+ +  E_- \hat{p}_- + \overrightarrow{p}_1 &=& 0
    \end{eqnarray}
 where the four vectors of the $\widetilde{\chi}_{1,2}^0$ in this frame are $(E_1,~
 \overrightarrow{p}_1)$ and $(m_i,~\overrightarrow{0})$, respectively. Equations (\ref{eq1})-(\ref{eq3}) give
 \begin{equation}\label{eq4}
    |  E_+ \hat{p}_+ +  E_- \hat{p}_- | = \frac{\sqrt{(M_{\ell^+ \ell^-}^2 - m_1^2-m_i^2)^2 - 4 m_1^2 m_i^2}}{2 m_i}
 \end{equation}
 which in the limit where $M_{\ell^+ \ell^-}$ attains its maximum ($= m_i - m_1$) reduces to
 \begin{equation}\label{eq5}
    |  E_+ \hat{p}_+ +  E_- \hat{p}_- |= 0
 \end{equation}
 further implying
 \begin{eqnarray} \label{eq6}
   E_\pm &=& \frac{m_i - m_1}{2} \\ \label{eq7}
   cos\theta &=& -1
 \end{eqnarray}
 i.e. the one constraint  (\ref{eq4}) is replaced by (\ref{eq5})-(\ref{eq7}) for events with maximal dilepton invariant mass.

 \subsection*{Spread of LSP mass measurement}
In the discussion of Section \ref{sec:zz} we showed that if we have the perfect event $M_{ee}= M_{\mu\mu} = M_{\ell \ell}^{max}$ then the extracted masses in (\ref{m1eqn}) both give $m_1' = m_1''= m_1$. In a finite-size event sample we can only expect to approach perfection, say $M_{ee,\mu\mu } =  M_{\ell \ell}^{max} - \epsilon$. In this case the Lorentz transforms defined in (\ref{betaeqn1}) will not give the LSP rest frames, where by definition their energies are
$E_{1,2} = m_1$, but rather frames where
\begin{equation}\label{psigma}
   E_{1,2} =  \frac{M_{ee,\mu\mu }^2 - m_1^2 - m_i^2}{2 m_i} \approx m_1 + \epsilon (1 - \frac{m_1}{m_i})
\end{equation}
as can be derived from (\ref{eq1})-(\ref{eq3}) above, with corresponding momenta
\begin{equation}\label{psigma}
   |\overrightarrow{p}_{1,2}| \approx \sqrt{2 \epsilon m_1(1 - \frac{m_1}{m_i})}
\end{equation}
Inverse-Lorentz-transforming \emph{these} four-vectors gives
\begin{equation}
\slashchar{{p}}_{1,2}^\mu =
\left(
  \begin{array}{cccc}
    \gamma & \beta_x \gamma & \beta_y \gamma & \beta_z \gamma \\
   \beta_x \gamma & 1 + (\gamma-1)\frac{\beta_x^2}{\beta^2} & (\gamma-1)\frac{\beta_x \beta_y}{\beta^2} & (\gamma-1)\frac{\beta_y\beta_z}{\beta^2} \\
    \beta_y \gamma & (\gamma-1)\frac{\beta_x \beta_y}{\beta^2} & 1+ (\gamma-1)\frac{\beta_y^2}{\beta^2} & (\gamma-1)\frac{\beta_y \beta_z}{\beta^2} \\
    \beta_z \gamma &(\gamma-1)\frac{\beta_x \beta_z}{\beta^2} &  (\gamma-1)\frac{\beta_y \beta_z}{\beta^2} & 1+ (\gamma-1)\frac{\beta_z^2}{\beta^2} \\
  \end{array}
\right)
\left(
  \begin{array}{c}
    m_1 + \epsilon (1 - \frac{m_1}{m_i}) \\
    O(\sqrt{ 2\epsilon m_1}) \\
    O(\sqrt{2\epsilon m_1}) \\
   O(\sqrt{ 2\epsilon m_1}) \\
  \end{array}
\right)
\end{equation}
for each $\overrightarrow{\beta} = \overrightarrow{\beta}_{1,2}$
which modifies (\ref{m1eqn}) to
 \begin{equation} \label{m1eqnnew}
  m_1' = \frac{\slashchar{\overrightarrow{p}}_x}{\beta_{1x} \gamma_1 + \beta_{2x} \gamma_2 \pm O(\sqrt{2\epsilon/m_1})} ~~~~~~~~~~~ m_1'' = \frac{\slashchar{\overrightarrow{p}}_y}{\beta_{1y} \gamma_1 + \beta_{2y} \gamma_2 \pm O(\sqrt{2\epsilon/m_1})}
\end{equation}
The precise spread in the distribution of values of $m_1$ therefore depends on a convolution of the $\widetilde{\chi}_{i}^0$-velocity-distribution and three-body phase space, but from the above we can already
 see it is roughly of order $\sqrt{2\epsilon m_1}$.


\begin{thebibliography}{99}


\bibitem{invmass}
H. Bachacou, I. Hinchliffe, and F. E. Paige,
``Measurements of masses in SUGRA models at LHC."
Phys.Rev.\textbf{D62}:015009 (2000);
E. Lytken,
 ``Derivation of some kinematical formulas in SUSY decay chains."
ATLAS note ATL-PHYS-COM-2004-001;
J.~M. Butterworth, J. Ellis and A.~R. Raklev,
``Reconstructing sparticle mass spectra using hadronic decays.''
JHEP \textbf{0705}, 033 (2007);
B.K. Gjelsten , D.J. Miller  , and P. Osland.
``Measurement of the gluino mass via cascade decays for SPS 1a."
JHEP \textbf{0506}:015 (2005);
P. Huang, N. Kersting, and H.H. Yang,
``Extracting MSSM Masses From Heavy Higgs Decays to Four Leptons at the LHC."
Phys. Rev. \textbf{D77}, 075011 (2008).

\bibitem{Kersting:wip}
N. Kersting, work in progress.

\bibitem{Barr}
A.~Barr, G.~G.~Ross, and M.~Serna,
``The Precision Determination of Invisible-Particle Masses at the LHC,"
Phys.Rev.\textbf{D78}, 056006 (2008).

\bibitem{benchmarks}
B.C.Allanach et al., ``The Snowmass points and slopes: Benchmarks for SUSY searches,"
Eur.Phys.J.\textbf{C25},113-123 (2002).

\bibitem{HERWIG65}
G.~Corcella {\it et al.},
JHEP {\bf 0101}: 010 (2001);
S.~Moretti, K.~Odagiri, P.~Richardson, M.H.~Seymour and B.R.~Webber,
JHEP {\bf 0204}: 028 (2002).

\bibitem{CTEQ6}
J. Pumplin {\it et al.} (CTEQ Collaboration),
JHEP \textbf{0207}:012 (2002);

D. Stump {\it et al.} (CTEQ Collaboration),
JHEP \textbf{0310}:046 (2003);

\bibitem{Bian}
G. Bian {\it et al.},
``Wedgebox analysis of four-lepton events
from neutralino pair production at the LHC,''
Eur.Phys.J.\textbf{C53}, 429 (2008).


\bibitem{Kersting:ss}  N.~Kersting, ``On Measuring Split-SUSY Gaugino Masses at the LHC,"
arXiv:0806.4238 [hep-ph]


\bibitem{Huang1}
P. Huang, N. Kersting and H.H. Yang,
``Extracting MSSM masses from heavy Higgs boson decays to four leptons
 at the CERN LHC,''
Phys.Rev.\textbf{D77}, 075011 (2008).

\bibitem{Bisset}
M.~Bisset, N.~Kersting, and R.~Lu, ``Improving SUSY Spectrum Determinations at the LHC with
Wedgebox and Hidden Threshold Techniques,"
arXiv:0806.2492 [hep-ph].


\bibitem{Kawagoe}
K.~Kawagoe, M.~M.~Nojiri and G.~Polesello,
``A new SUSY mass reconstruction method at the CERN LHC,''
Phys.Rev.\textbf{D71}, 035008 (2005).

\bibitem{Cheng}
H.-C. Cheng {\it et al.},
``Mass determination in SUSY-like events with missing energy,''
JHEP \textbf{0712}, 076 (2007).


\bibitem{mT2}
C.~G.~Lester and D.~J.~Summers,
``Measuring masses of semi-invisibly decaying particles
 pair produced at hadron colliders,''
Phys.Lett.\textbf{B463}, 99 (1999);
A.~Barr, C.~Lester and P.~Stephens,
``m(T2): The truth behind the glamour,''
J. Phys.\textbf{G29}, 2343 (2003).

\bibitem{mtvar}
W.~S.~Cho et al.,
``Measuring superparticle masses at hadron collider
using the transverse mass kink,''
JHEP {\bf 0802}, 035 (2008),
Phys. Rev. Lett. {\bf 100}, 171801 (2008);
A.~J.~Barr, B.~Griparios  and C.~G.~Lester,
``Weighing WIMPS with kinks at colliders:  invisible particle mass
 measurements from endpoints,''
JHEP {\bf 0802}, 014 (2008);
M.~M.~Nojiri et al.,
``Inclusive transverse mass analysis for squark and gluino mass
 determination,''
 JHEP {\bf 0806}, 035 (2008).



\bibitem{Gjelsten}
B.~K.~Gjelsten, D.~J.~Miller, and P.~Osland,
``Measurement of SUSY Masses via Cascade Decays
for SPS 1a,"
JHEP {\bf 0506}, 015 (2005).


\end{thebibliography}
\end{document}